\documentclass[12pt]{article}
\usepackage[utf8]{inputenc}
\usepackage{geometry}
\usepackage{graphicx}
\usepackage{amsmath, amssymb}
\usepackage{chemformula} 
\usepackage{authblk}
\usepackage[numbers,sort&compress]{natbib}
\usepackage{setspace}

\title{Fate of amine-based selenenyl sulfides during interaction with glutathione reductase: a molecular dynamics perspective}
\author[1,2]{Vishnu Rama Chari}
\author[1]{Raghu Nath Behera\thanks{Corresponding author: rbehera@goa.bits-pilani.ac.i}}
\affil[1]{Department of Chemistry, Birla Institute of Technology \& Science, Pilani, K K Birla Goa Campus, Zuarinagar, Sancoale, Goa 403726, India}
\affil[2]{School of Chemical Sciences, Goa University, Taleigao Plateau, Goa 403206, India}

\date{}

\begin{document}
\onehalfspacing	
	\maketitle
	
\begin{abstract}
In the search for small organoselenium-based mimics of the glutathione peroxidase (GPx) enzyme, it has been observed that selenenyl sulfides (RSeSG) derived from amine-based GPx mimics have the potential to be reduced at the catalytic site of glutathione reductase (GR), thereby enhancing the catalytic efficiency of these mimics in biological systems. However, molecular insights into these interactions are lacking due to the absence of force field parameters for \ch{Se-S} containing compounds. In this study, we present force field parameters for selenenyl sulfides with a phenyl selenide backbone developed using the General Amber Force Field (GAFF). Employing these parameters, a 200 ns molecular dynamics (MD) simulation of RSeSG was performed. The results indicate that both the amino nitrogen and its substituent significantly influence the interaction of RSeSG at the catalytic site of GR.
\end{abstract}
	

\section{Introduction}
\begin{figure}[h!]
	\centering
	\includegraphics[width=0.6\textwidth]{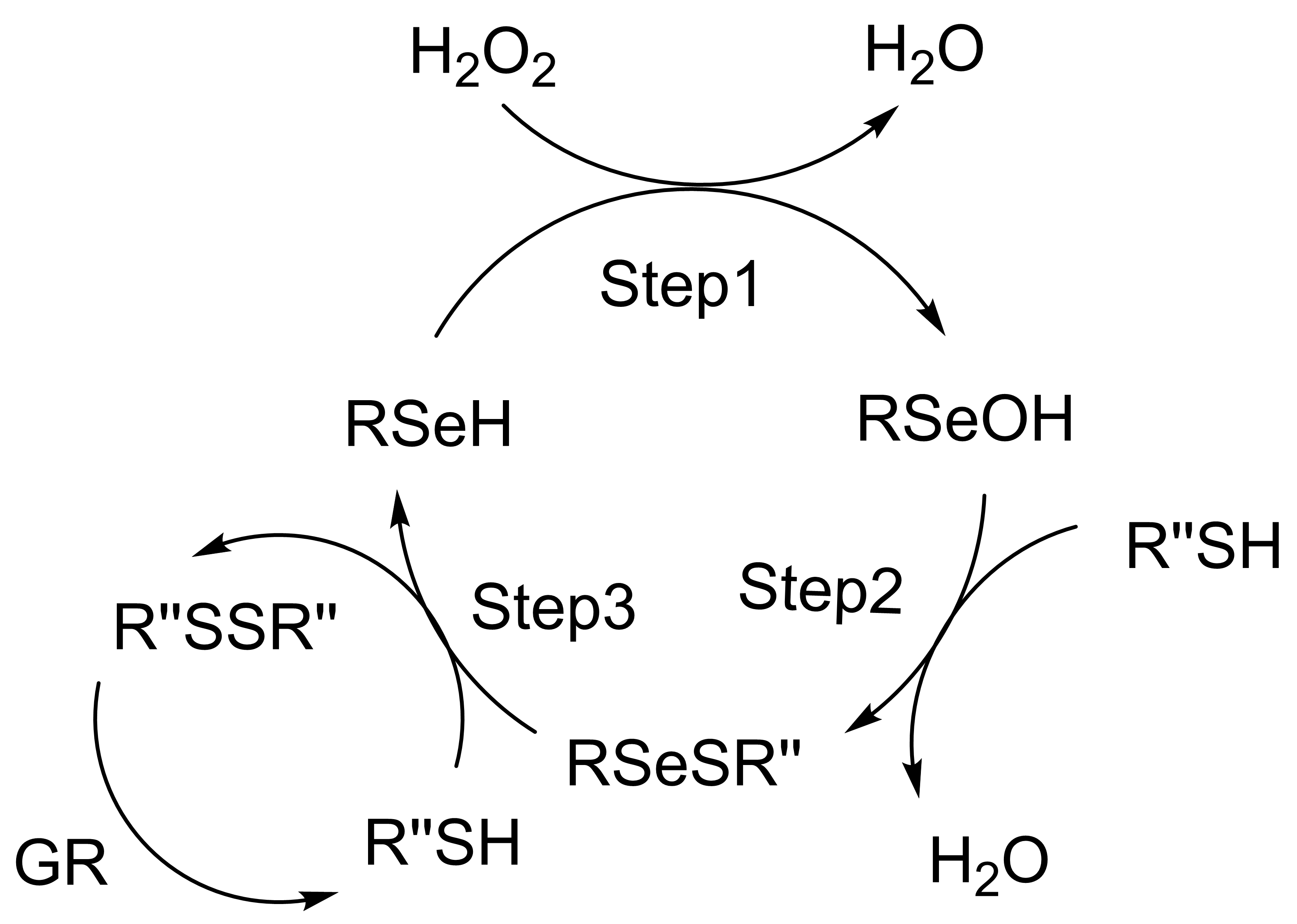}
	\caption{The catalytic cycle of GPx mimics in biological assay. Disulfides (GSSG) generated during the catalytic cycle are converted to glutathione (GSH) using glutathione reductase (GR) enzyme.}  
	\label{catal}
\end{figure}

Glutathione peroxidase (GPx) enzyme mitigates oxidative stress by reducing harmful reactive oxygen species from the cell. GPx contains selenocysteine (Sec) instead of cysteine (Cys) at its active center~\cite{flohe1973glutathione}. The active site in GPx enzyme is characterized by a catalytic triad formed by the selenium atom from Sec45, two nitrogen atoms from glutamine (Gln80), and tryptophan (Trp158)~\cite{bhowmick2015introduction}. Many small organoselenium compounds possessing such catalytic triads have been utilized as GPx mimics~\cite{bhabak2010functional,bhowmick2015insights,barbosa2017organoselenium}. The GPx mimics follow a catalytic cycle akin to that of the native enzyme~\cite{iwaoka1994model,bhowmick2015insights}. As illustrated in Figure \ref{catal}, in the initial step of the catalytic cycle, the selenol moiety (RSeH) is oxidized to selenenic acid moiety (RSeOH) by reducing harmful peroxides. Subsequently, this oxidized RSeOH is regenerated back to its RSeH form through reactions: first, it reacts with one molecule of thiol (R''SH), forming a selenenyl sulfide (RSeSR''), followed by the reaction of RSeSR'' with a second R''SH to produce a disulfide (R''SSR'') and regenerate RSeH.

For native enzymes, glutathione (GSH) functions as the thiol cofactor, and during its catalytic cycle, the disulfide (GSSG) is regenerated by glutathione reductase (GR). Given that selenium often behaves as a congener of sulfur, previous studies \cite{giles2012hydrogen,yu2021new} have employed biological assays to investigate the involvement of interaction of RSeSR'' ($i.e.$ RSeSG) at the active site of GR. Preliminary findings in literature \cite{giles2012hydrogen,yu2021new,kumar2023introduction} suggests a potential interaction of RSeSG at the active site of GR. Yu et al.~\cite{yu2021new}, utilizing molecular docking studies, have demonstrated that the enhanced catalytic activity of certain GPx mimics is due to their interaction with the catalytic site of GR. While molecular docking offer insights into static protein-ligand interactions, it lacks the capability to depict the dynamic structural changes resulting from these interactions. For this reason, MD simulations are often favored. Despite the potential of MD to simulate selenium-containing compounds, there are relatively few studies documented in the literature, partly due to the scarcity of appropriate force field parameters~\cite{torsello2016general,pedron2023novel,fellowes2022simulating,loschwitz2021dataset}, which limits the availability of MD simulation results.

In one of our recent works \cite{chari2024can}, employing density functional theory (DFT) and molecular docking studies, we showed that amine-based RSeSGs have potential to be reduced at the catalytic site of GR. This suggests that similar to the regeneration of GSH from GSSG at the GR's catalytic site, RSeSG can regenerate a RSeH and a GSH molecule, thereby enhancing biological catalytic activity. However, due to the unavailability of force field parameters for organoselenium compounds containing \ch{Se-S} bond, these interactions could not be modeled. Therefore, in this work, the force field parameters for organoselenium compounds containing \ch{Se-S} bond were generated using generalized amber force field (GAFF) principle ~\cite{wang2004development,case2023ambertools}. Later, these parameters were used for MD simulation to understand the interactions between the RSeSG and GR. Selenenyl sulfides belonging to phenyl selenides, di-2(N-cyclohexyl,N-(methylamino)-methyl)phenyl diselenide (CMP) and  N, N-dialkyl benzylamine-based diselenides (DMP) were chosen for this study. The study demonstrates that the presence of the amino group and its substituent in phenyl selenide-based GPx mimics are crucial for their interaction with the active site of GR, enabling the molecule to properly orient itself for reduction at the catalytic site.

\section{Computational methods}

Geometry of all the compounds were initially optimized at B3LYP/cc-pVTZ level of theory. Frequency calculation was performed to verify that the obtained structure has no imaginary frequencies, which indicated that the optimized geometry is a minimum on the potential energy surface. The opt=modredundant keyword was employed to allow selective modification of internal coordinates during optimization. Specific bond lengths, bond angles, or dihedral angles were constrained and scanned. This approach enabled the exploration of potential energy surfaces and the evaluation of structural changes while maintaining control over key geometrical parameters.

To prepare initial structures for molecular dynamics (MD) simulations, the RSeSG ligand was docked into the active site of GR (PDB ID: 1GRA) using AutoDock Vina, following established protocols\cite{trott2010autodock}. The receptor and ligand were prepared in PDBQT format, and docking was performed using an optimized grid box centered on the active site. Multiple binding poses were generated, and the pose with the lowest predicted binding affinity was selected for further study. Similarly, the initial structure of flavin adenine dinucleotide (FAD) were obtained in its binding site as GR is a flavoprotein and FAD plays a role as a prosthetic groups in stabilizing GR~\cite{krauth1985fad}. The geometry of the best-docked RSeSG pose was used to compute the molecular electrostatic potential (ESP) at the B3LYP/6-31G(d) level of theory with Gaussian. Implicit solvent effects were included via the SMD model with water as the solvent. The Gaussian output containing ESP data was provided as input to ANTECHAMBER (AmberTools)~\cite{case2023ambertools} for Restrained Electrostatic Potential (RESP) charge fitting. This procedure produced AMBER-compatible mol2 files and since ANTECHAMBER does not support Se atom, workaround is used by replacing Se with S~\cite{loschwitz2021dataset}. The resulting files were used for subsequent force field preparation and molecular dynamics setup. The details of these parameters are provided in the supplementary information.

MD simulations were performed using GROMACS 2024 suite of programs~\cite{abraham2015gromacs}. The ligand parameterization was performed using GAFF as implemented via AmberTools for all ligands, while the AMBER03 force field was applied for the protein. One molecule of model compound, test compound or protein-ligand complex was solvated in a cubic box of 12.0 \AA{} with TIP3P water~\cite{jorgensen1983comparison}. The simulation protocol comprised energy minimization, equilibration and production phase. The temperature was regulated using velocity-rescale thermostat with a coupling time constant of 0.1 ps. Following 5~ns NVT equilibration, an 5~ns NPT equilibration step was conducted. Pressure coupling was achieved using the Parrinello-Rahman barostat with isotropic coupling. The temperature coupling was maintained as in the NVT step. Production MD simulations were performed in the NPT ensemble using the leap-frog integrator with a 2 fs timestep for a total of 200-500 ns. All bonds to hydrogen atoms were constrained using the LINCS algorithm. Non-bonded interactions were treated using a Verlet cutoff scheme with neighbor searching on a grid, applying a cutoff of 1.5 nm for both van der Waals and electrostatic interactions.

\section{Results and Discussion}

\begin{figure}[ht]
	\includegraphics[width=1\textwidth]{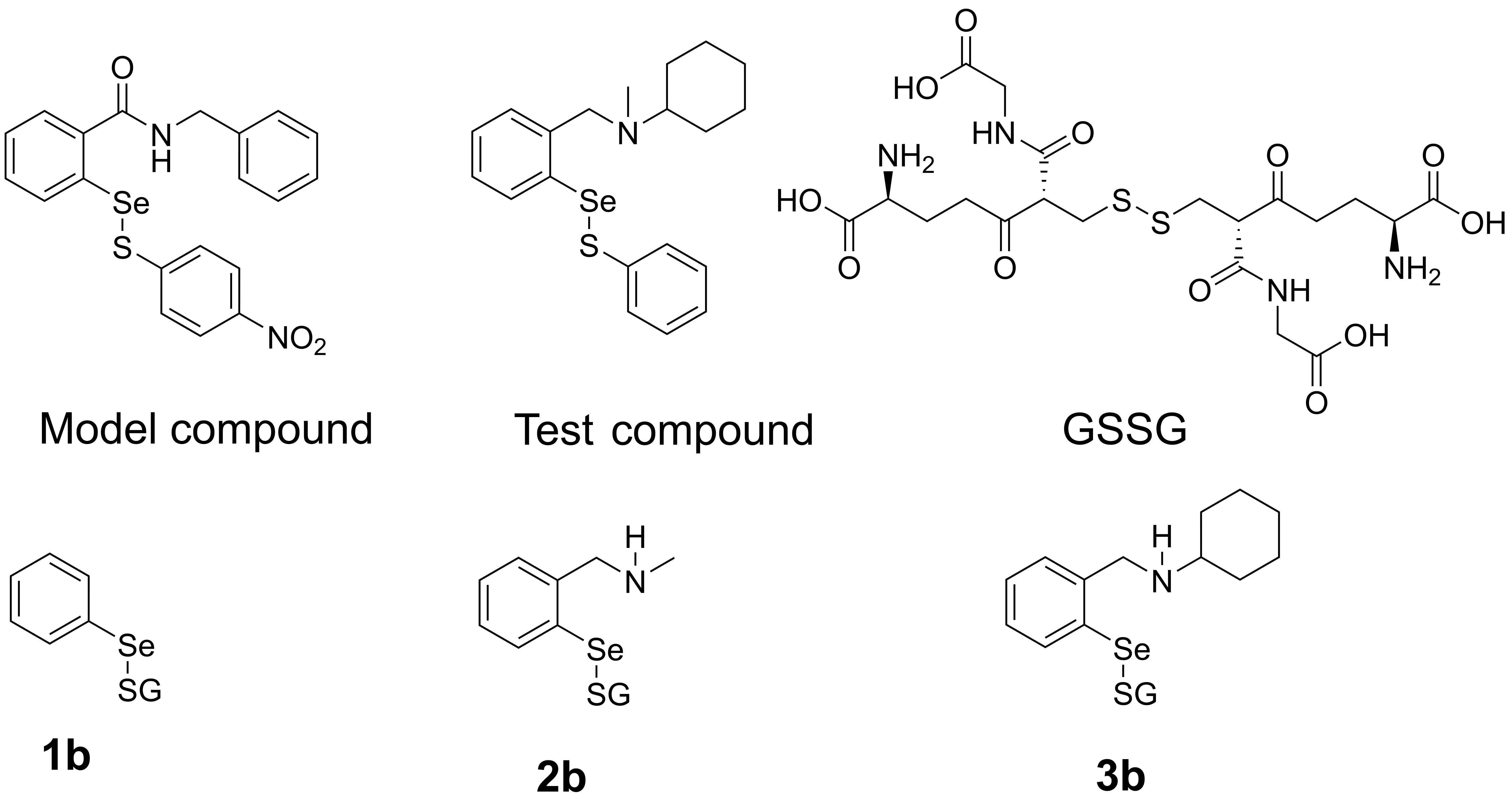}
	\caption{Structures of the compounds studied.}
	\label{structures}
\end{figure}

Investigating the reactions of type 
\begin{equation}
	\ch{RSeSG} + \ch{GR}^{SH}_{SH} \rightleftharpoons \ch{RSeH} +\ch{GR}^{S}_{S}]
	\label{reduction}
\end{equation}
indicated that \ch{RSeSG} has a potential to be reduced at the catalytic site of \ch{GR}~\cite{chari2024can}.  Among the compounds studied, amine-based \ch{RSeSG}s derived from compounds with phenyl selenide backbone exhibited more positive redox potentials compared to \ch{GSSG}, favoring reduction at the \ch{GR} active site. This trend was further supported by molecular docking analyses. Given the inherently dynamic nature of biological processes, we aimed to validate these findings through MD simulations. Since no established force field exists for \ch{Se-S} containing compounds, we derived GAFF-based parameters and applied them to run MD simulations for the \ch{RSeSG-GR} complexes. The structures of the compounds under study are given in Figure \ref{structures}. The results of these simulations are discussed in the following sections.

\subsection{GAFF parameters for RSeSG}
GAFF Hamiltonian~\cite{wang2004development} (Eq. \ref{gaff}) has been used to generate the force field parameters for RSeSG. 
 
\begin{multline}
	E_{pair} = \sum_{\text{bonds}} K_r (r - r_{eq})^2 + \sum_{\text{angles}} K_{\theta} (\theta - \theta_{eq})^2 \\
	+ \sum_{\text{dihedrals}} \frac{V_n}{2} \left[ 1 + \cos(n \phi - \gamma) \right]
	+ \sum_{i<j} \left[ \frac{A_{ij}}{R_{ij}^{12}} - \frac{B_{ij}}{R_{ij}^6} - \frac{q_i q_j}{\epsilon R_{ij}} \right]
	\label{gaff}
\end{multline}

Where, r$_{eq}$ and $\theta_{eq}$ are equilibrium structural parameters; K$_{r}$, K$_{\theta}$, V$_{n}$ are force constants; $n$ is multiplicity and $\gamma$ is phase angle for torsional angle parameters.

There are limited number of crystal structures for selenenyl sulfides containing phenyl selenide backbone, we chose N-benzyl-2-{[(4-nitrophenyl)sulfanyl]selanyl}benzamide (model compound) (CCDC deposition no. 2166327)~\cite{shang2023facile} for generating the force field parameters for RSeSG. The flexibility and thus the conformational space sampled by a molecule is affected by the dihedral parameters. We used fourier fit approach to optimize the associated GAFF dihedral force constants and phase parameters. The phase parameters were then converted to either 0.0 or 180.0 as per GAFF ideology. The fitted parameters for each compound are given in the Table S1-S4 and Figure S1-S6. A systematic comparison of key geometric parameters obtained for the model compound was performed using three approaches $viz.$ the  direct DFT optimization, fitting DFT-derived energies to the GAFF Hamiltonian, and analysis of trajectories generated from a 250 ns MD simulations of the ligand in water as solvent (see Table \ref{compare}). Specifically, the geometric parameters obtained via DFT represent the gas-phase equilibrium structure at the quantum level, while the fitted values reflect the minimum-energy geometry derived from GAFF functional forms using quantum energetic benchmarks. The parameters obtained from MD simulations denote ensemble-averaged values and their fluctuations under the GAFF force field with explicit solvent and thermal sampling over 250 ns.

\begin{table}[h!]
	\centering
	\caption{Geometric parameters obtained for the model compound with DFT optimized geometry, fitting DFT energetic benchmark to GAFF Hamiltonian,  and 250 ns MD simulation using GAFF derived parameters. }
	\label{compare}
	\begin{tabular}{lrrr}
		\hline
		Geometric 		& DFT & Fitted from&Obtained from \\
		parameter\textsuperscript{a} & & DFT Energies &MD simulation\\
		\hline
		Se$-$S 		& 2.252		& 2.284 	& 2.301 $\pm$ 0.079\\
		C1$-$Se 	& 1.952		& 1.968 	& 2.006 $\pm$ 0.059\\
		C4-C1-Se	& 121.657	& 119.132 	& 120.097 $\pm$ 2.611\\
		C3-C1-Se	& 120.752	& 121.977 	& 120.308 $\pm$ 2.644\\
		C1-Se-S		& 100.764	& 101.363 	& 104.792 $\pm$ 3.515\\
		C2-S-Se		& 115.786	& 107.550 	& 112.449 $\pm$ 4.278\\
		\hline
	\end{tabular} \\
	\small{\textsuperscript{a}Distances in \AA{} and angles in $\deg$.}
\end{table}

The data presented in Table \ref{compare} show that bond lengths and angles obtained by fitting DFT energies to the GAFF Hamiltonian generally fall between the DFT-optimized values and the ensemble-averaged values from MD simulations. A slight deviation observed in the bond length and bond angles obtained from MD simulation is possibly due to the effect of dynamic sampling and thermal fluctuations. The standard deviations indicate the magnitude of these fluctuations. Overall, this comparison demonstrates that the developed GAFF parameters reasonably capture both quantum mechanical minima and the distribution of structural parameters under dynamic conditions.

\subsection{Validation of the derived GAFF parameters}

To assess the accuracy of the derived GAFF parameters, we selected the selenenyl sulfide derivative of CMP with PhSH (referred to as the test compound). The experimental $^{77}$Se NMR chemical shift for test compound reported in the literature is 574 ppm\cite{iwaoka1994model}.  $^{77}$Se NMR shifts were computed for the test compound relative to dimethyl selenide in methanol.  Our calculations, based on 1000 structures extracted from a 500 ns MD simulation of test compound in water, using B3LYP/6-31G(d,p) level of theory show an average $\delta$ value of 686 ppm while using B3LYP/6-31++G(2df,2pd) level of theory show an average $\delta$ value of 761 ppm  with respect to dimethyl selenide, indicating a deviation of 112 and 187 ppm respectively compared to the literature value. Several factors may contribute to this deviation from the experimental value some of which could be due to (i) not refining the extracted structures at QM level, (ii) not using explicit solvent molecules during QM calculation (iii) large range of chemical shift ($\sim$3000 ppm) for Se nucleus, and (iv) sensitivity of 
$^{77}$Se NMR shift to the orientation of neighboring chemical groups~\cite{torsello2016general}. Despite this deviation, the results demonstrate reasonable agreement, validating the applicability of the derived GAFF parameters for modeling these selenenyl sulfide species.

\subsection{Testing of the derived GAFF parameters}
Earlier using density functional theory, for the reaction given in Eq. \ref{reduction} we have shown that RSeSGs: \textbf{1b}, \textbf{2b}, and \textbf{3b} belonging to amine-based GPx mimics had shown relatively positive redox potential (-0.251, -0.252 and -0.257 V respectively) compared to that of GSSG (E\textsubscript{GSSG|GSH} = -0.289 V)~\cite{chari2024can}. Relatively positive redox potential suggests that \ch{RSeSG} can get easily reduced at the catalytic site of GR compared to GSSG. However, this reaction can take place only if \ch{RSeSG} can approach the catalytic site of GR. To test this possibility, the derived GAFF parameters were employed for understanding the interaction of three different RSeSG at the active site of GR (PDB ID: 1GRA)~\cite{karplus1989substrate}. 1GRA corresponds to the crystal structure of GR from human erythrocytes in which the active site is bound to GSSG, and hence chosen for the study.

For reduction of GSSG at the active site of GR $\left( \ch{GSSG} + \ch{GR}^{SH}_{SH} \rightleftharpoons \ch{GSH} +\ch{GR}^{S}_{S}] \right)$, S\textsubscript{GSSH} interacts with the S\textsubscript{Cys-58}. The distance of interaction measured in 1GRA~\cite{karplus1989substrate} between the S\textsubscript{Cys-58} and S1\textsubscript{GSSH} is 4.23~\AA{} and that of S\textsubscript{Cys-58} and S2\textsubscript{GSSH} is 5.87~\AA{}. MD simulation between 1GRA and GSSG was run as a benchmark calculation and the average distance between S\textsubscript{Cys-58}-S1\textsubscript{GSSH} and S\textsubscript{Cys-58}-S2\textsubscript{GSSH} are found to be 4.58 and 5.99~\AA{}, respectively (deviation from experiment is 0.35 and 0.12~\AA{}, respectively). This indicated that the 200 ns MD simulation of GSSG with 1GRA could be used to study the approach of GSSG at the active site of GR for reduction.

To assess the conformational stability of the molecular dynamics simulations, several structural parameters were calculated: root mean square deviation (RMSD) of ligand heavy atoms relative to the protein backbone, radius of gyration for protein backbone and ligand, solvent-accessible surface area (SASA) of the protein-ligand complex, and root mean square fluctuation (RMSF) of the protein backbone and ligand. These results are summarized in Table S5.

\begin{figure}
	\centering
	\begin{tabular}{cc}
		\includegraphics[width=0.45\textwidth]{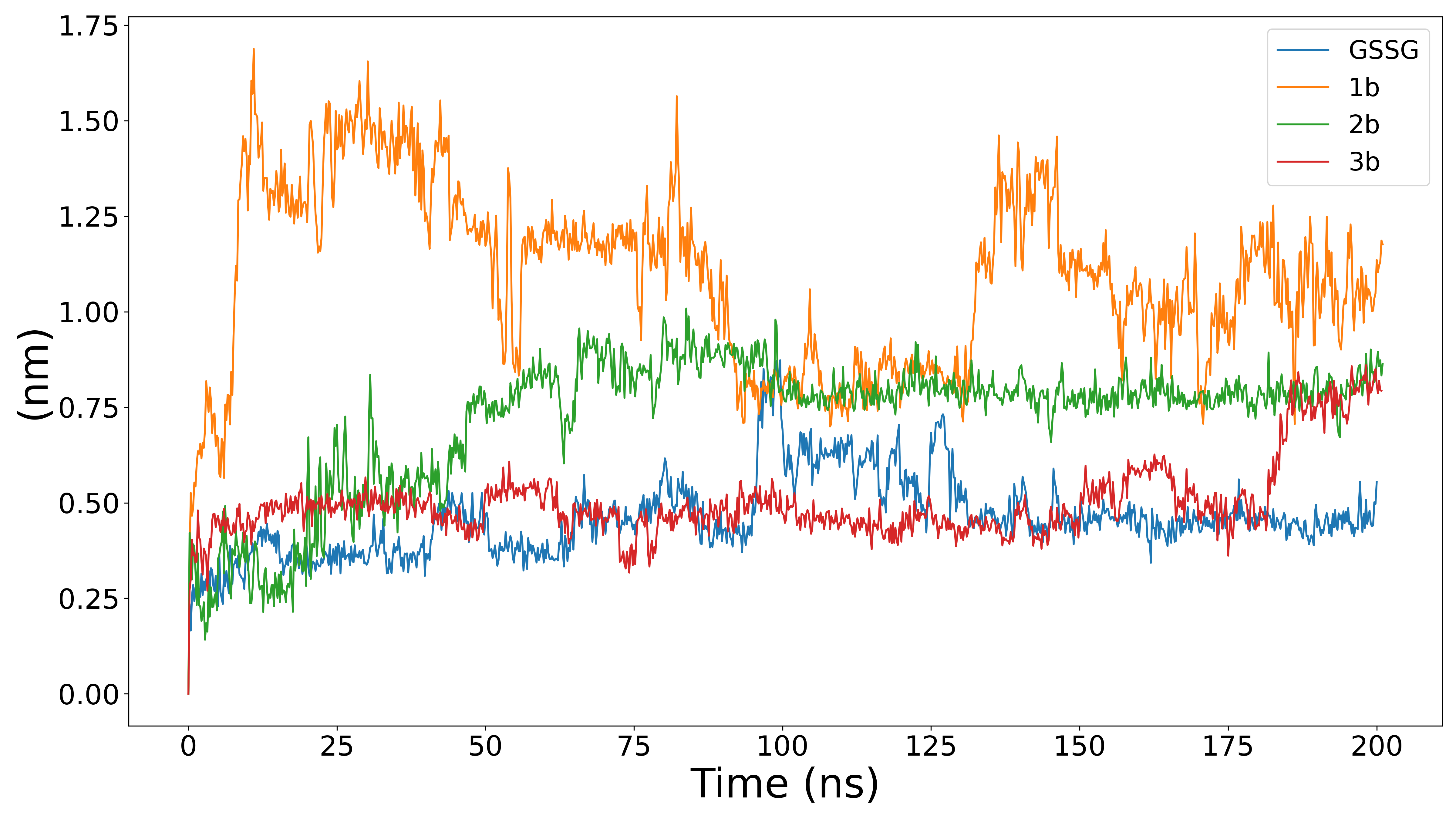}&
		\includegraphics[width=0.45\textwidth]{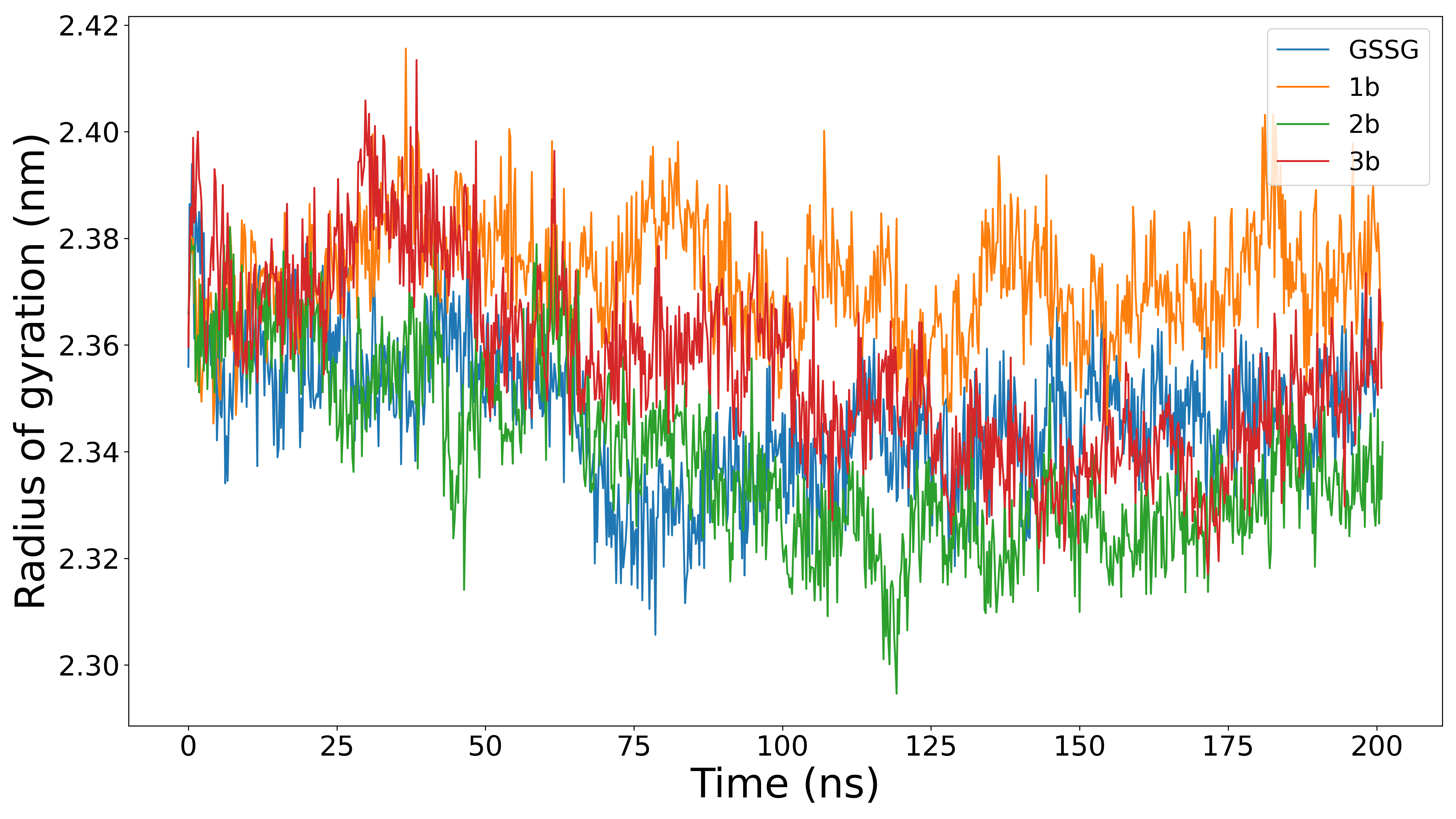}\\
		\includegraphics[width=0.45\textwidth]{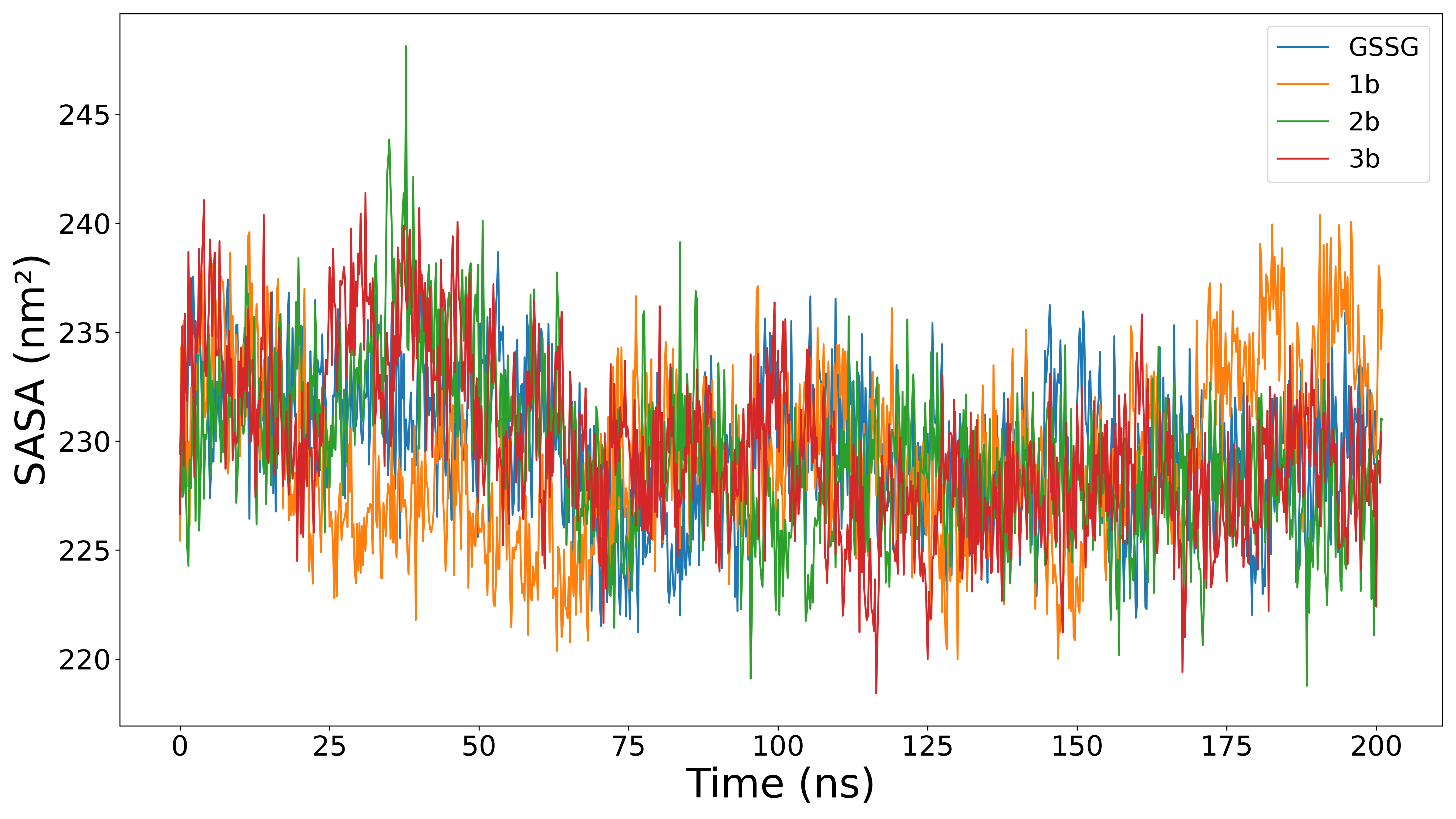}	&
		\includegraphics[width=0.45\textwidth]{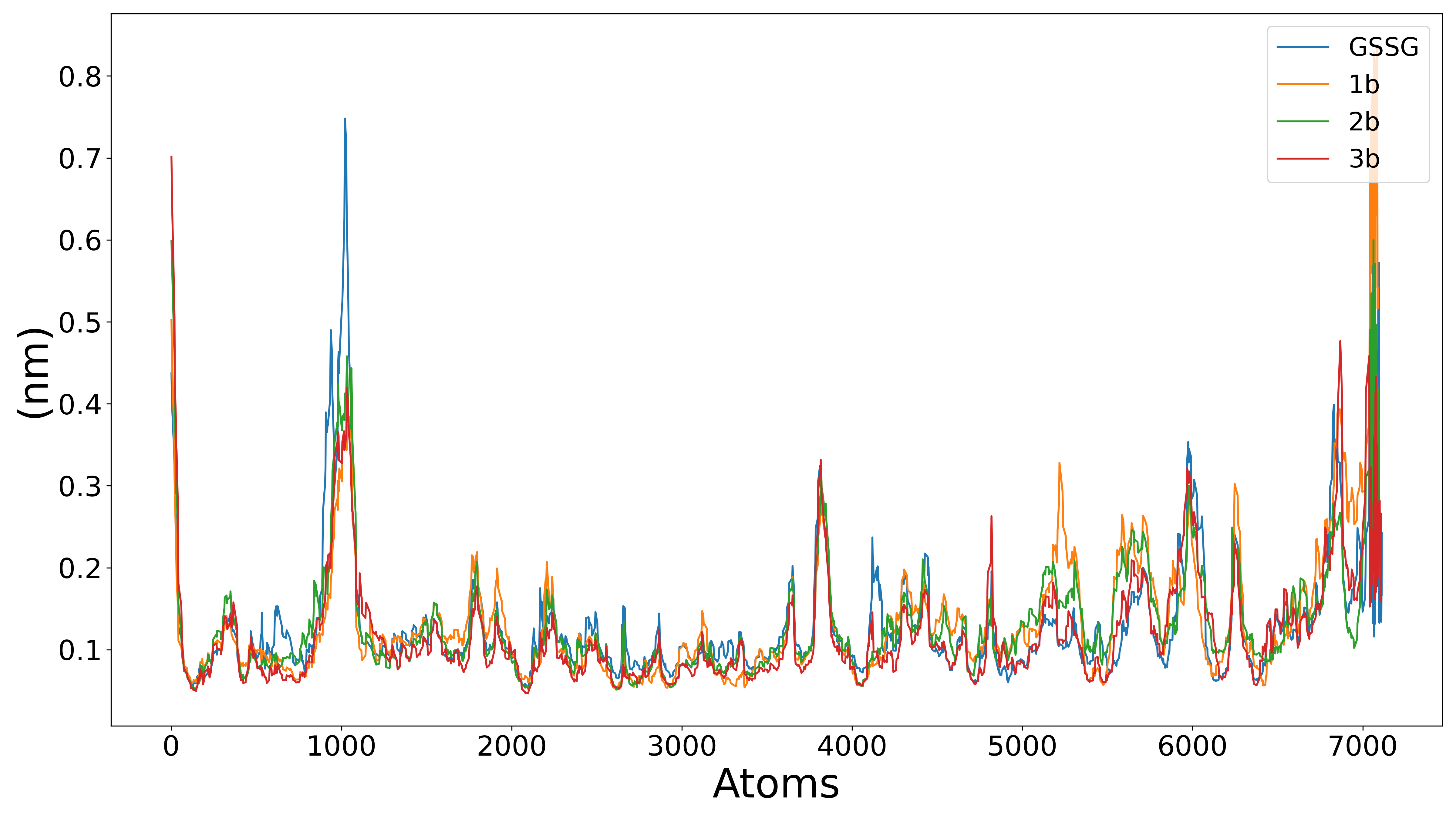}\\	
		
	\end{tabular}
	\caption{Heavy-atom RMSD of the ligand relative to protein backbone (top left), radius of gyration for protein backbone and ligand  (top right), solvent-accessible surface area (SASA) of the protein–ligand complex (bottom left), and RMSF of both protein backbone and ligand (bottom right).}
	\label{analysis}
\end{figure}

Analysis of Figure \ref{analysis}  and Figure S7 reveals that RMSD values follow a similar trend to the distance between Se/S\textsubscript{Lig} and S\textsubscript{Cys-58}. This correlation between ligand-protein distance and RMSD indicates that structural changes in the complexes are primarily driven by ligand-protein interactions. Representative conformations for each protein-ligand complex are displayed in Figure S8.

The average RMSD values are 1.09, 0.72, 0.50, and 0.46 nm for compounds \textbf{1b}, \textbf{2b}, \textbf{3b}, and GSSG, respectively. The high RMSD of 1.09 nm and longer distances between the Se\textsubscript{\textbf{1b}} and S\textsubscript{Cys58} reflects the inability of \textbf{1b} to form a stable complex with the active site of the protein. \textbf{2b} is seen to form a stable complex at the active site of the catalyst thereby showing the possibility to get catalyzed at the active site of GR.  On the other hand, \textbf{3b} forms a stable complex in the catalytic pocket but remains at a distance away from the  S\textsubscript{Cys-58} for any possible reaction to take place.

Analysis of the MD trajectory (as shown in Supplimentary videos) aimed at elucidating the origin of selectivity for amino-substituted ligands reveals that Tyr-114 and Ser-30 play pivotal roles. Tyr-114 is known to form hydrogen bonds with the backbone nitrogen of Gly and with the nitrogen of the Cys residue in GSSG \cite{francescutti1996peroxynitrite}. In the GSSG:GR complex, the approach distance between S\textsubscript{Lig} and S\textsubscript{Cys-58} varies in a correlated manner with two key interactions: (i) the distance between N\textsubscript{Lig} of the Cys residue in GSH within GSSG and O\textsubscript{Tyr-114}, and (ii) the distance between N\textsubscript{Lig} of the Cys residue in GSH within GSSG and O\textsubscript{Ser-30}. The RMSD values for these correlated distances are 1.83 and 1.75 \AA, respectively (Figure S9). In compounds \textbf{2b} and \textbf{3b}, the amino group nitrogen replaces the Cys nitrogen of GSSG in interactions with Tyr-114 and Ser-30.

It is observed from Figure S7 that in the presence of amino group, Se of selenenyl sulfides can come in proximity to S\textsubscript{Cys-58}. Compared to GSSG, \textbf{2b} with methyl as amino substituent approaches the S\textsubscript{Cys-58} at an average distance of 5.07 \AA{}, thereby showing a possibility of getting catalysed at the active site of GR. 
During the final 50 ns of MD simulation, the distance between N of the amino group on aromatic ring in \textbf{2b} and O\textsubscript{Ser-30} as well as the N\textsubscript{Cys} of GSSG and O\textsubscript{Ser-30} remains around $\sim$4 \AA{} (Figure S9), which is consistent with long hydrogen bond formation~\cite{baker1984hydrogen,francescutti1996peroxynitrite}. The corresponding distances  between N of the amino group on aromatic ring in \textbf{2b} and O\textsubscript{Tyr-114} is  $\sim$4 \AA{} and for N\textsubscript{Cys} of GSSG are in range of $\sim$6 \AA{}. These interactions in \textbf{2b}, thus orient the ligand favorably for interaction with S\textsubscript{Cys-58} of GR.

However, in \textbf{3b} (with cyclohexyl ring as amino substituent), average distance between Se\textsubscript{\textbf{3b}} and S\textsubscript{Cys-58}  during the MD simulation is 6.89 \AA{}. In contrast, compound \textbf{3b} exhibits distances of $\sim$8 \AA{} and $\sim$4 \AA{} between N of the amino group and O\textsubscript{Tyr-114}, and  O\textsubscript{Ser-30} respectively, indicating that the bulky cyclohexyl substituent prevents effective interaction with Tyr-114, thereby hindering proper orientation for catalysis at the GR active site.

\textbf{1b} which contains only phenyl selenide backbone (absence of amino group) during most of the simulation time  remains away from S\textsubscript{Cys-58} (as far as $\sim$18 \AA). The average distance between Se\textsubscript{\textbf{1b}} and S\textsubscript{Cys-58} is 12.23 \AA. However, during $\sim$ 90 to 130 ns the average distance between Se\textsubscript{\textbf{1b}} and S\textsubscript{Cys-58} drops to  $\sim$6.5 \AA{} and again it moves away from  S\textsubscript{Cys-58}. The distance between N\textsubscript{Cys} belonging to GSH of \textbf{1b} and O\textsubscript{Tyr-114}, and N\textsubscript{Cys} belonging to GSH of \textbf{1b} and O\textsubscript{Ser-30} also follow a similar trend as seen from Figure S9 (b). Due to absence of the amino nitrogen in \textbf{1b} these interactions takes place at the GSH part rather than at the aromatic ring of \textbf{1b}. Distances exceed 10 \AA{}, indicating that these interactions do not contribute to orienting \textbf{1b} in a manner that allows Se to interact with S\textsubscript{Cys-58}. 

Therefore even though \textbf{1b} and \textbf{2b} show similar redox potential (-0.251 and -0.252 V)~\cite{chari2024can}  for getting reduced at the catalytic site of GR, \textbf{2b} due to the presence of less bulkier amino substituent can approach the active site of GR and get reduced, showing higher GPx-like catalytic activity. Thus, we can summarize that the distance of approach between Se\textsubscript{Lig} and S\textsubscript{Cys-58}  is primarily governed by the bulkiness of the amino substituent and highlights the importance of nature of amino group in such GPx mimics.

The radius of gyration and SASA do not vary much for the different ligands and are about 2.3 nm and 229 nm$^2$, respectively.  Values closer to that of GSSG indicates that the ligands are likely to have a similar conformational stability and solvent exposure as that of the natural substrate within the active site of GR. The RMSF for all the other ligands are similar to GSSG showing that the protein has similar flexible regions during the simulation indicating that the RSeSG binding doesn't alter the protein structure differently than GSSG binding.

\section{Conclusion}

Due to unavailability of  force field parameters for compounds containing \ch{Se-S} bonds we first generated GAFF parameters for seleneyl sulfides of GPx mimics containing phenyl selenide backbone. The RMSD, RMSF, Rg and SASA analysis show that except for \textbf{1b} (selenenyl sulfide of only phenyl selenide) these compounds form stable complex at the active site of GR during the 200 ns MD simulation. Furthermore, it was noticed that the presence of amino group is essential for phenyl selenide based ligands to approach the active site of GR. Effective interaction between Tyr-114 and Ser-30 are found to orient the ligand suitably towards the active site of GR. The bulky cyclohexyl group present in \textbf{3b} increases the distance between Tyr-114 and the amino nitrogen of \textbf{3b}, thereby preventing the Se atom of \textbf{3b} from interacting with the active site. In contrast, compound \textbf{2b}, which features a methyl group on the amino nitrogen, is able to interact with the S\textsubscript{Cys-58} of the active site. This is enabled by the favorable orientation resulting from interactions of the amino nitrogen of \textbf{2b} with Tyr-114 and Ser-30. Since \textbf{1b} lacks an amino nitrogen, it is unable to approach the active site, and as a result, remains uncatalyzed. These observations highlight the crucial role of the amino group and its interactions with neighboring residues in facilitating the proper binding and orientation of RSeSG at the catalytic site of GR. In continuation of our earlier study~\cite{chari2024can}, The present results indicate that the knowledge of redox potential and favourable interactions as seen in the molecular docking study alone provides an incomplete picture of the interaction between 
RSeSG and GR. Importantly, this molecular dynamics study shows that there is a possibility for RSeSG to get reduced at the active site of GR.

\section*{Acknowledgments}
We acknowledge the DST-FIST supported computational facility to the Department of Chemistry, and the HPC facility of Birla Institute of Technology and Science Pilani, K. K. Birla Goa Campus for this work. VRC is thankful to UGC-SAP and DST-FIST funding to School of Chemical Sciences, Goa University.
	
\bibliographystyle{WileyNJD-ACS}
\bibliography{ref}
 	
\end{document}